\documentstyle[12pt,amssymb,epsfig]{article}

\begin{document}

\newcommand{\dh}{\eth}

\title{Characteristic Surface Data for the Eikonal Equation}
\author{Ezra T. Newman and Alejandro Perez \\
Dept. of Physics and Astronomy, \\
University of Pittsburgh, Pgh., PA 15260, USA}
\date{Sept. 8, 1998 }
\maketitle

\begin{abstract}
A method of solving the eikonal equation, in either flat or curved
space-times, with arbitrary Cauchy data, is extended to the case of data
given on a characteristic surface. We find a beautiful relationship between
the Cauchy and characteristic data for the same solution, namely they are
related by a Legendre transformation. From the resulting solutions, we study
and describe the wave-front singularities that are associated with their
level surfaces (the characteristic surfaces or ``big wave-fronts'').
\end{abstract}

\section{Introduction}

The high frequency limit of the wave equation is given by the eikonal
equation, written in an arbitrary space-time as:{\ 
\begin{equation}
g^{ab}({x^{a}})\partial _{a}S({x^{a}})\partial _{b}S({x^{a}})=0  \label{eiko}
\end{equation}
}where the{\ $x^{a}=(x^{i},t)$ }are any local coordinates, and{\ \ $g^{ab}({%
\ x^{a}})$ }is the metric of the given space-time. The level surfaces of a
solution of Eq{.(\ref{eiko}), $S({x^{a}})=const.$, (which need not be smooth
every place and could have self-intersections) }are three dimensional
characteristic surfaces (the ``big wave-fronts'' in the terminology of Arnold%
\cite{Arnold}), and the sections {$t=constant$ }of these surfaces are the
two-dimensional (``small'') wave fronts. The vector field {$%
l^{a}=g^{ab}\partial _{b}S$ }is tangent to the null geodesic that generate
the characteristic surfaces.

In flat space-time the eikonal equation becomes{\ 
\begin{equation}
\eta ^{ab}\partial _{a}S\partial _{b}S={(\partial _{t}S)}^{2}-{(\partial
_{x}S)}^{2}-{(\partial _{y}S)}^{2}-{(\partial _{z}S)}^{2}=0.  \label{eik*}
\end{equation}
}

In Sec.2 we review the method \cite{gil} to give a general solution of the
eikonal equation in flat space-time adapted to appropriate Cauchy data given
at $t=t_{0}$. In Sec.3 we modify the method so that the eikonal equation is
solved with arbitrary \textit{characteristic} data given at null infinity.
The relation between both methods is studied in Sec.4 where we find that the
Cauchy and characteristic data are related by a Legendre transformation. The
wave-front singularities of the level surfaces of the resulting solutions
are described parametrically in Sec.5 and finally the generalization of our
results to asymptotically flat spaces-times is given in Sec.6. \ 

\section{Solutions of the Eikonal Equation}

>From the point of view of the theory of partial differential equation the
eikonal equation is a homogeneous first order non-linear partial
differential equation; there exist a solution $S^{*}$, called `complete
integral', depending on three arbitrary constants\cite{landau}, e.g., in
flat space-time the function 
\begin{equation}
{S^{*}(x^{i},t,{\alpha }_{i})=x^{i}{\alpha }^{i}-t\sum {({\alpha }_{i}}}%
\noindent )^{2}  \label{*}
\end{equation}
is easily seen to satisfy (\ref{eik*}). \\

\noindent
{\bf Remark 1.}
{\it The fact that the equation is homogeneous plays no role in this section,
but will be crucial in the generalization to characteristic data.} \\

\noindent
{\bf Remark 2.}
{\it From now on we will treat the problem of the eikonal in flat space; we leave
the discussion of general space-times to the end.}\\

It is possible to generate a ``general integral'', i.e., a solution of the
eikonal equation depending on an arbitrary function, by means of the
following procedure: First, define the function $S^{**}(x^{i},t,{\alpha }
_{i})$ of the coordinates and the free parameters ${\alpha }_{i}$ as

\begin{equation}
S^{**}(x^{i},t,{\alpha }_{i})\equiv S^{*}(x^{i},t,{\alpha }_{i})-H({\alpha }
_{i}),  \label{palito}
\end{equation}

\noindent where $H({\alpha }_{i})$ is any function of the ${\alpha }_{i}$'s.

Next, think of the $\alpha _{i}$'s as functions of the space-time points
obtained from the following conditions:

\begin{equation}
\partial S^{**}/\partial \alpha _{i}=\partial S^{*}/\partial \alpha _{i}-
\partial H/\partial \alpha _{i}=0  \label{lucia}
\end{equation}

\noindent Equation (\ref{lucia}) determine the space-time dependence of the $%
\alpha _{i}$'s ($\alpha _{i}=A_{i}(x^{i},t)$) given that

{\ 
\begin{equation}
\left| {\frac{{\partial ^{2}S^{**}(x^{i},t,{\alpha }_{i})}}{{\partial \alpha
_{i}\partial \alpha _{j}}}}\right| \neq 0.  \label{vani}
\end{equation}
}This condition can fail in lower dimensional regions called the caustics.
This issue will be returned to in Sec.5.

Substituting $\alpha _{i}=A_{i}(x^{i},t)$ into equation (\ref{palito}) we
eliminate the $\alpha _{i}$, and obtain

\begin{equation}
S^{**}(x^{i},t)=S^{*}(x^{i},t,A_{i}(x^{i},t))-H(A_{i}(x^{i},t)).
\label{pooh}
\end{equation}

\noindent It is easy to verify that, because of the condition (\ref{lucia}),

\begin{equation}
\partial _{a}S^{**}=\partial _{a}S^{*}
\end{equation}

\noindent which means that $S^{**}(x^{i},t)$ is a new solution of the
eikonal equation (\ref{eiko}) determined by an arbitrary function $H$. We
can determine the free function $H$ so that the solution (\ref{pooh})
satisfies initial Cauchy data at $t=t_{0}$. We denote the Cauchy data by $%
S_{Cauchy}(x^{i})$. Conditions (\ref{lucia}) imply that at $t=t_{0},$ $%
\alpha _{i}=\partial S_{Cauchy}/\partial x^{i}$. Inverting these relations
we obtain $x^{i}=X^{i}(\alpha _{i})$, and replacing them in (\ref{pooh}) at $%
t=t_{0}$ we find the sought for relation:

{\ 
\begin{equation}
H(\alpha _{i})=S^{*}(X^{i}(\alpha _{i}),t_{0},\alpha
_{i})-S_{Cauchy}(X^{i}(\alpha _{i})).  \label{a}
\end{equation}
}

This last equation relates the arbitrary function $H(\alpha _{i})$ with the
Cauchy data, $S_{Cauchy}(x^{i}),$ at $t=t_{0}$ and allows us to construct
solutions of the eikonal equation in flat space-time for any initial data. 
\newline
\qquad We now change the set of the $\alpha ^{i}$'s for new parameters that
are more appropriate to the study of asymptotically flat spaces and we
rewrite our previous equations in terms of them. A complete integral of the
eikonal equation, equation (1), can be written in terms of new parameters $%
(\beta ,\zeta ,\bar{\zeta})$ as:

\begin{equation}
S^{*}(x^{a},\beta ,\zeta ,\bar{\zeta})=\beta x^{a}l_{a}(\zeta ,\bar{\zeta}).
\label{b}
\end{equation}

\noindent where

\begin{equation}
l_{a}(\zeta ,\bar{\zeta})={\frac{1}{\sqrt{2}(1+\zeta \bar{\zeta})}}((1+\zeta 
\bar{\zeta}),(\zeta +\bar{\zeta}),-i(\zeta -\bar{\zeta}),(\zeta \bar{\zeta}%
-1))  \label{1*}
\end{equation}

\noindent is the null covector pointing in the $(\zeta ,\bar{\zeta})$
direction. The $(\zeta ,\bar{\zeta})$ are the stereographic coordinates that
parametrize the sphere of null directions.

>From (\ref{b}) and (\ref{1*}), we get the relations between the new
parameters $(\beta ,\zeta ,\bar{\zeta})$, and the old $\alpha $'s:

\begin{eqnarray}
\alpha _{1} &=&{\frac{\beta }{\sqrt{2}}}{\frac{\zeta +\bar{\zeta}}{1+\zeta 
\bar{\zeta}}}  \nonumber \\
\alpha _{2} &=&-i{\frac{\beta }{\sqrt{2}}}{\frac{\zeta -\bar{\zeta}}{1+\zeta 
\bar{\zeta}}}  \nonumber \\
\alpha _{3} &=&{\frac{\beta }{\sqrt{2}}}{\frac{\zeta \bar{\zeta}-1}{1+\zeta 
\bar{\zeta}}},  \label{cordichange}
\end{eqnarray}

\noindent and

\begin{equation}
\beta =\sqrt{2\sum \alpha _{i}^{2}}
\end{equation}

In terms of the new parameters Eq.(\ref{palito}) reads:

\begin{equation}
S^{**}(x^{a},\beta ,\zeta ,\bar{\zeta})=\beta x^{a}l_{a}(\zeta ,\bar{\zeta}
)-H(\beta ,\zeta ,\bar{\zeta}),  \label{beta}
\end{equation}

\noindent where $H(\beta ,\zeta ,\bar{\zeta})$ is an arbitrary function that
will be determined by the initial conditions. Conditions (\ref{lucia}) on ($%
\beta ,\zeta $ ,$\bar{\zeta})$ become

\begin{eqnarray}
x^{a}l_{a}(\zeta ,\bar{\zeta})-{\frac{\partial H}{\partial {\beta }}}(\beta
,\zeta ,\bar{\zeta})|_{\zeta ,\bar{\zeta}} &=&0  \label{three} \\
\beta x^{a}m_{a}(\zeta ,\bar{\zeta})-{\ \eth H}(\beta ,\zeta ,\bar{\zeta}%
)|_{\beta } &=&0  \label{three*} \\
\beta x^{a}\bar{m}_{a}(\zeta ,\bar{\zeta})-\bar{{\ \eth}}H(\beta ,\zeta ,%
\bar{\zeta})|_{\beta } &=&0.  \label{three**}
\end{eqnarray}

\noindent
{\bf Remark 3.}
{\it We have replaced the derivatives with respect to $\zeta $ and $\bar{\zeta}$
respectively by 
\begin{eqnarray*}
{\eth} &=&(1+\zeta \bar{\zeta})\frac{\partial }{\partial \zeta }, \\
\bar{{\ \eth}} &=&(1+\zeta \bar{\zeta})\frac{\partial }{\partial \overline{%
\zeta }}
\end{eqnarray*}
and used the fact that $\dh l_{a}(\zeta ,\bar{\zeta})=m_{a}$ and ${\eth}%
\bar{{\ \eth}}l_{a}=n_{a}-l_{a}$ where ($l_{a},n_{a},m_{a},\overline{m}_{a})$
form a null Minkowski space tetrad for each ($\zeta ,\bar{\zeta}).$} \\

The function $H(\beta ,\zeta ,\bar{\zeta})$ can be determined by means of
the same procedure using the conditions $\alpha _{i}=\partial S_{Cauchy}/
\partial x^{i}$ at $t=t_{0}$ and relations (\ref{cordichange}) to obtain the 
$x^{i}=X^{i}(\beta ,\zeta ,\bar{\zeta})$, and finally rewriting (\ref{a})

\[
H(\beta ,\zeta ,\bar{\zeta})=S^{*}(X^{i}(\beta ,\zeta ,\bar{\zeta}
),t_{0},\beta ,\zeta ,\bar{\zeta})-S_{Cauchy}(X^{i}(\beta ,\zeta ,\bar{\zeta}
)). 
\]

\section{Characteristic Data for the Eikonal Equation\qquad}

The eikonal equation, being hyperbolic, admits a characteristic formulation.
Even though the results of this section can be applied to any characteristic
hypersurface in Minkowski, in flat space (as in any asymptotically simple
space-time) there are two preferred characteristic surfaces, namely future
and past null infinity respectively. In the following we will formulate the
characteristic problem in terms of data given at future null infinity, $%
\frak{{I}^{+}}$ . $\frak{{I}^{+}}$ has the topology of $S^{2}\times R$ ; we
choose Bondi coordinates on it, namely ($\zeta $, $\bar{\zeta})$ on the $%
S^{2}$ and the retarded time $u_{B}$ along $R$. In an analogous manner as
for the Cauchy problem, the characteristic data at future null infinity will
be defined by a function of ($u_{B},\zeta $, $\bar{\zeta}$ );

\begin{equation}
S_{Characteristic}=L(u_{B},\zeta ,\bar{\zeta}).  \label{cara}
\end{equation}

The goal of this section is to develop a method to construct solutions of
the eikonal equation geometrically adapted to the characteristic data, Eq.( 
\ref{cara}), at $\frak{{I}^{+}}$.

\noindent
{\bf Remark 4.}
{\it In asymptotically flat space-times, in the neighborhood of future null
infinity, $\frak{I}^{+},$ there is a preferred class of coordinates referred
to as Bondi coordinates. Given a Bondi system $(u_{B},\zeta ,\bar{\zeta})$
at $\frak{I}^{+},$ a new system\cite{new} $(u,\zeta ,\bar{\zeta})$ can be
defined by $(u,\zeta ,\bar{\zeta})=(L(u_{B},\zeta ,\bar{\zeta}),\zeta ,\bar{%
\zeta}).$ The characteristic data, (\ref{cara}), can be thought of as being
generated by this coordinate change representing a one parameter family of
arbitrary $u=const.$ slices of $\frak{I}^{+}.$} \\

In flat space-times we can define a two parameters family of null surfaces
by:

\begin{equation}
S^{*}=x^{a}l_{a}(\zeta ,\bar{\zeta})  \label{cono}
\end{equation}

As was pointed out in \cite{gil} Eq.(\ref{cono}) has a dual interpretation.
For ($\zeta ,\bar{\zeta})$ kept constant, its level surfaces define null
planes intersecting the time axis at a time equal the value of $S^{*}$ and
with its direction given by $(\zeta ,\bar{\zeta})$; on the other hand, for a
fixed value of $x^{a}$ it represents the light cone cut at $\frak{{I}^{+}}$
of the space time point $x^{a}$ in the interior, i.e., it represents the
intersection of the null cone from $x^{a}$ with $\frak{{I}^{+}}$.

We can think of the characteristic data (\ref{cara}) geometrically, as
defining a one parameter family of cuts at $\frak{{I}^{+}}$ in terms of $%
u=L(u_{B},\zeta ,\bar{\zeta})=const.$ (It is assumed that this can be
inverted so that the cuts are given by $u_{B}=L^{-1}(u=const,\zeta ,\bar{%
\zeta}).)$ With this point of view, we construct a solution of the eikonal
equation (\ref{eiko}) (corresponding to the characteristic data $%
L(u_{B},\zeta ,\bar{\zeta})$), such that the family of null surfaces in the
interior are defined by the null geodesics normal to the family of cuts at
infinity given by $L(u_{B},\zeta ,\bar{\zeta})=const.$. In order to do so we
will generalize the method of section II.

Defining

\begin{equation}
S^{**}(x^{a},\zeta ,\bar{\zeta})=L(S^{*},\zeta ,\bar{\zeta}%
)=L(x^{a}l_{a}(\zeta ,\bar{\zeta}),\zeta ,\bar{\zeta}),  \label{c}
\end{equation}
we see immediately that it is a solution of the eikonal equation depending
on two free parameters. [Note the duplication of notation which arises from
the different meanings to the same object; $S^{*}=u_{B}=x^{a}l_{a}(\zeta ,%
\bar{\zeta}).]$ By putting the requirement on $\zeta $ and $\bar{\zeta}$
that ${\eth}S^{**}=\bar{{\ \eth}}S^{**}=0$, i.e.,{\ 
\begin{eqnarray}
\dot{L}(S^{*},\zeta ,\bar{\zeta})x^{a}m_{a}(\zeta ,\bar{\zeta})+{\eth}%
L(S^{*},\zeta ,\bar{\zeta}) &=&0  \nonumber \\
\dot{L}(S^{*},\zeta ,\bar{\zeta})x^{a}\bar{m}_{a}(\zeta ,\bar{\zeta})+\bar{{%
\ \eth}}L(S^{*},\zeta ,\bar{\zeta}) &=&0,  \label{micon}
\end{eqnarray}
}where $\dot{L}\equiv \partial _{S^{*}}L$, we can solve for $\zeta $ and $%
\bar{\zeta}$ in terms of $x^{a}$, i.e., equations (\ref{micon}) give us 
\begin{equation}
\zeta =\Gamma (x^{a}) { and }\bar{\zeta}=\overline{\Gamma }(x^{a})
\end{equation}
except at the caustics when\cite{Arnold2} 
\begin{equation}
\left| 
\begin{array}{cc}
{\eth}^{2}S^{**} & {\eth}\bar{{\ \eth}}S^{**} \\ 
{\eth}\bar{{\ \eth}}S^{**} & \bar{{\ \eth}}^{2}S^{**}
\end{array}
\right| =0.
\end{equation}
This issue will be discussed in Sec.5.

Finally replacing ($\zeta ,\bar{\zeta})$ in (\ref{c}) and differentiating we
find

\begin{equation}
\partial _{a}S^{**}(x^{a},\Gamma (x^{a}),\overline{\Gamma }(x^{a}))=\dot{L}
\;l_{a}(\Gamma (x^{a}),\overline{\Gamma }(x^{a})).
\end{equation}

Therefore, the function

\begin{equation}
S^{**}(x^{a},\Gamma (x^{a}),\overline{\Gamma }(x^{a}))  \label{cha}
\end{equation}

\noindent satisfies the eikonal equation, and by construction (\ref{c}) it
is adapted to the characteristic data defined by the function $L(u_{B},\zeta
,\bar{\zeta})$ at $\frak{{I}^{+}}$. The null normals to level surfaces 
\begin{equation}
S^{**}(x^{a},\Gamma (x^{a}),\overline{\Gamma }(x^{a})=const.
\end{equation}
are normal to the cuts $L(u_{B},\zeta ,\bar{\zeta})=const.$ at $\frak{{I}%
^{+}.}$ Note that the fact that $S^{**}(x^{a})$ is a new solution of the
eikonal equation is a consequence of property of the eikonal equation of
being homogeneous in $\partial _{a}S$.

\section{Relation between the Cauchy and Characteristic Constructions}

\noindent \qquad In this section we give the connection between the two
methods of construction. Earlier we showed how to relate the Cauchy data, $%
S_{Cauchy}(x^{i}),$ with the arbitrary function $H(\beta ,\zeta ,\bar{\zeta}%
) $ of Sec.2, so that any solution of the eikonal equation can be cast in
the form of equation (\ref{beta}). Therefore, there must be a relationship
of the characteristic construction to the construction via Cauchy data and
hence a relationship between the functions $L(x^{a}l_{a}(\zeta ,\bar{\zeta}%
),\zeta ,\bar{\zeta})$ and $H(\beta ,\zeta ,\bar{\zeta})$.

We first note that though in both methods there is an arbitrary function of
three variables, in the characteristic method there appear only two
parameters ($\zeta ,\bar{\zeta})$ while in the Cauchy method there are the
three ($\beta ,\zeta ,\bar{\zeta}).$

We can reduce the three to two by solving Eq.(\ref{three})

\begin{equation}
x^{a}l_{a}(\zeta ,\bar{\zeta})-{\frac{\partial H}{\partial \beta }}(\beta
,\zeta ,\bar{\zeta})=0,  \label{cero}
\end{equation}

\noindent for $\beta =\beta (x^{a}l_{a},\zeta $ ,$\bar{\zeta})$ or changing
notation and using $u_{B}=x^{a}l_{a}$, we have $\beta =\beta (u_{B},\zeta ,%
\bar{\zeta})$. Now thinking of (\ref{cero}) as an implicit relation defining 
\textit{either }$u_{B}$ = $U(\beta ,\zeta ,\bar{\zeta})\equiv $ ${\frac{%
\partial H}{\partial \beta }}(\beta ,\zeta ,\bar{\zeta})$ or $\beta $ = $%
\beta (u_{B},\zeta ,\bar{\zeta})$. Note that if we treat $H$ as a function
of $(u_{B},\zeta ,\bar{\zeta}),$i.e., $H=H(\beta (u_{B},\zeta ,\bar{\zeta}%
),\zeta ,\bar{\zeta})$ then \newline
\begin{equation}
\dot{H}\equiv \partial _{u_{B}}H|_{\zeta ,\bar{\zeta}}={\frac{\partial H}{%
\partial \beta }}(\beta ,\zeta ,\bar{\zeta})\dot{\beta}
\end{equation}
or\qquad 
\begin{equation}
\dot{\beta}=\frac{\dot{H}}{\frac{\partial H}{\partial \beta }}  \label{2*}
\end{equation}

We replace $\beta =\beta (u_{B},\zeta ,\bar{\zeta})$ into the two
conditions, Eqs.(\ref{three*}) and (\ref{three**}), obtaining

\begin{eqnarray}
\beta (u_{B},\zeta ,\bar{\zeta})x^{a}m_{a}(\zeta ,\bar{\zeta})-{\eth }
H(\beta (u_{B},\zeta ,\bar{\zeta}),\zeta ,\bar{\zeta})|_{\beta } &=&0 
\nonumber \\
\beta (u_{B},\zeta ,\bar{\zeta})x^{a}\bar{m}_{a}(\zeta ,\bar{\zeta})-\bar{{\
\eth }}H(\beta (u_{B},\zeta ,\bar{\zeta}),\zeta ,\bar{\zeta})|_{\beta } &=&0,
\label{uno}
\end{eqnarray}

\noindent which appear similar to Eqs.(\ref{micon}), namely:

\begin{eqnarray}
\dot{L}(u_{B},\zeta ,\bar{\zeta})x^{a}m_{a}(\zeta ,\bar{\zeta})+{\eth}%
L(u_{B},\zeta ,\bar{\zeta})|_{u_{B}} &=&0  \nonumber \\
\dot{L}(u_{B},\zeta ,\bar{\zeta})x^{a}\bar{m}_{a}(\zeta ,\bar{\zeta})+\bar{{%
\ \eth}}L(u_{B},\zeta ,\bar{\zeta})|_{u_{B}} &=&0.  \label{pito}
\end{eqnarray}

We explicitly write $|_{\beta }$ and $|_{u_{B}}$ in the $\eth$ -operators to
mean that the angular derivatives are taken keeping $\beta $ or $u_{B}$
constant respectively; also $\dot{L}$ means $\partial _{u_{B}}L|_{\zeta ,%
\bar{\zeta}}$ for any $L(u_{B},\zeta ,\bar{\zeta})$.

\noindent
{\bf Remark 5}
{\it As we mentioned earlier, Eqs.(\ref{uno}) or (\ref{pito}) implicitly define $%
\zeta =\Gamma (x^{a})$ and $\bar{\zeta}=\overline{\Gamma }(x^{a})$
everywhere except at the caustics. They can be approached in a limiting
fashion.} \\

Given an arbitrary function $F(\beta ,\zeta ,\bar{\zeta})$ and $\beta
(u_{B},\zeta ,\bar{\zeta})$ there is the following relation between
differential operators.

\begin{eqnarray}
{\eth }F(\beta ,\zeta ,\bar{\zeta})|_{\beta } &=&{\eth }F(\beta ,\zeta ,\bar{
\zeta})|_{u_{B}}-(\partial F/\partial \beta )(\beta ,\zeta , \bar{\zeta}){\
\eth }\beta |_{u_{B}}  \nonumber \\
\bar{{\eth }}F(\beta ,\zeta ,\bar{\zeta})|_{\beta } &=&\bar{{\eth }} F(\beta
,\zeta ,\bar{\zeta})|_{u_{B}}-(\partial F/\partial \beta )(\beta ,\zeta ,%
\bar{\zeta})\bar{{\eth }}\beta |_{u_{B}}.  \label{bu}
\end{eqnarray}

Using these relations to replace the $\eth$ and $\bar{{\eth}}$ derivative
operators at $\beta $ constant by operators at $u_{B}$ constant in (\ref
{pito} ) we obtain:

\begin{eqnarray}
\beta x^{a}m_{a}(\zeta ,\bar{\zeta})-{\eth }H(\beta ,\zeta ,\bar{\zeta}
)|_{u_{B}}+(\partial H/\partial \beta )(\beta ,\zeta ,\bar{\zeta}){\eth }
\beta |_{u_{B}} &=&0  \nonumber \\
\beta x^{a}m_{a}(\zeta ,\bar{\zeta})-\bar{{\eth }}H(\beta ,\zeta ,\bar{ \zeta%
})|_{u_{B}}+(\partial H/\partial \beta )(\beta ,\zeta ,\bar{\zeta})\bar{{\
\eth }}\beta |_{u_{B}} &=&0.  \label{tres}
\end{eqnarray}

Applying relations (\ref{bu}) to the function $F=u_{B}=x^{a}l_{a}(\zeta ,%
\bar{\zeta})$, thought of as $u_{B}=U(\beta ,\zeta ,\bar{\zeta})$, via the
following steps;\qquad 
\begin{equation}
{\eth}F(\beta ,\zeta ,\bar{\zeta})|_{\beta }={\eth}(x^{a}l_{a})=x^{a}m_{a}
\end{equation}

\begin{equation}
{\eth }F(\beta ,\zeta ,\bar{\zeta})|_{u}=\eth u|_{u}=0
\end{equation}

\begin{equation}
(\partial F/\partial \beta ){\eth }\beta |_{u_{B}}=(\partial u/\partial
\beta ){\eth }\beta |_{u_{B}}=\dot{\beta}^{-1}{\eth }\beta |_{u_{B}}
\end{equation}
we get the following important equation:

\begin{equation}
\dot{\beta}x^{a}m_{a}=-{\ \eth }\beta |_{u_{B}}.
\end{equation}

Finally inserting this relation, with Eq.(\ref{2*}), into (\ref{tres}) we
obtain

\begin{eqnarray}
\beta x^{a}m_{a}(\zeta ,\bar{\zeta})-{\eth }H(\beta ,\zeta ,\bar{\zeta}
)|_{u_{B}}-\dot{H}x^{a}m_{a}(\zeta ,\bar{\zeta}) &=&0  \nonumber \\
\beta x^{a}m_{a}(\zeta ,\bar{\zeta})-\bar{{\eth }}H(\beta ,\zeta ,\bar{ \zeta%
})|_{u_{B}}-\dot{H}x^{a}\bar{m}_{a}(\zeta ,\bar{\zeta}) &=&0,
\end{eqnarray}

\noindent which can be rewritten as

\begin{eqnarray}
{\eth }(u_{B}\beta -H)|_{u_{B}}+\dot{(u_{B}\beta -H)}x^{a}m_{a}(\zeta , \bar{
\zeta}) &=&0  \label{3*} \\
\bar{{\eth }}(u_{B}\beta -H)|_{u_{B}}+\dot{(u_{B}\beta -H)}x^{a}\bar{m}
_{a}(\zeta ,\bar{\zeta}) &=&0.  \nonumber
\end{eqnarray}

\noindent Comparing Eq.(\ref{3*}) with (\ref{pito}) we see that they are
identical when we set

\begin{equation}
L(u_{B},\zeta ,\bar{\zeta})=u_{B}\beta (u_{B},\zeta ,\bar{\zeta})-H{\large (}
\beta (u_{B},\zeta ,\bar{\zeta}),\zeta ,\bar{\zeta}{\large )}.  \label{4*}
\end{equation}

\noindent From equation (\ref{cero}) we also have that

\begin{eqnarray}
u_{B} &=&\partial H/\partial \beta ,  \label{5*} \\
\beta &=&\partial L/\partial u_{B}.  \nonumber
\end{eqnarray}

We see that the two data functions $L(u_{B},\zeta ,\bar{\zeta})$ and $%
H(\beta ,\zeta ,\bar{\zeta})$ are related by the Legendre transformation,
Eqs.(\ref{4*}) and (\ref{5*}). We have finally arrived at a very simple and
beautiful relation between the two methods. An essential property of the
eikonal equation for this relationship is that it is homogeneous in $%
\partial _{a}S$.

\section{Parametric Description of the Wave-Fronts}

Using the methods described above we can construct a general solution of the
eikonal equation either for the Cauchy or the (corresponding) characteristic
data. In the last section we showed that they are simply related by a
Legendre transformation in the variables $u_{B}$ and $\beta $. Once we have
the solution of the eikonal equation we can study the geometry of its
wave-fronts. In particular we are interested in the description of the
singularities developed by them, namely, its caustics.\newline

\qquad A key step in both methods consists of expressing the free
parameters, e.g., $(\beta ,\zeta $, $\bar{\zeta})$, contained in the
formalism as functions of the space-time points, $x^{a}$. In many cases the
problem of inverting Eq.(\ref{three}-\ref{three**}) or Eq.(\ref{micon}) in
order to get either ($\zeta $,$\bar{\zeta})$ or ($\beta ,\zeta $, $\bar{\zeta%
})$ as functions of the space-time coordinates $x^{a},$ can be a formidable
task and at times impossible. However, it is not absolutely necessary, since
it is possible to give a parametric description of the null surfaces defined
by Eq.(\ref{beta}) or Eq.(\ref{cha}) respectively (In the sequel we follow
the path given in ref.\cite{gil} for the case of the stationary eikonal
equation). \newline

In the Cauchy case we have three parameters ($\zeta ,\bar{\zeta}$,$\beta )$
in the initial data for the eikonal equation. See Eq.(\ref{beta}). We
introduce the new parameter $r$ together with ($\zeta $, $\bar{\zeta}$,$%
\beta )$ by means of the following equation;

\begin{equation}
r=\beta ^{-1}\eth\bar{{\eth}}S^{**}=x^{a}(n_{a}-l_{a})-\beta ^{-1}\eth\bar{{%
\eth}}H(\beta ,\zeta ,\bar{\zeta})|_{\beta }  \label{para**}
\end{equation}
and the previous equations;

\begin{eqnarray}
x^{a}l_{a}(\zeta ,\bar{\zeta})-{\frac{\partial H}{\partial {\beta }}}(\beta
,\zeta ,\bar{\zeta})|_{\zeta ,\bar{\zeta}} &=&0,  \label{para*} \\
\beta x^{a}m_{a}(\zeta ,\bar{\zeta})-{\eth }H(\beta ,\zeta ,\bar{\zeta})
&=&0,  \nonumber \\
\beta x^{a}\bar{m}_{a}(\zeta ,\bar{\zeta})-\bar{{\eth }}H(\beta ,\zeta , 
\bar{\zeta}) &=&0,  \nonumber
\end{eqnarray}

The four Eqs.(\ref{para*}) and (\ref{para**}) can be solved for the
coordinates $x^{a}$ in terms of ($\beta ,r,\zeta $,$\bar{\zeta}$), using the
orthonormality of the null tetrad:

\begin{equation}
x^{a}={\frac{\partial H}{\partial \beta }}(l^{a}+n^{a})-(r-{\frac{{\eth } 
\bar{{\eth }}H}{\beta }})l^{a}-{\frac{\bar{{\eth }}H}{\beta }}m^{a}-{\ \frac{%
{\eth }H}{\beta }}\bar{m}^{a}.  \label{betun}
\end{equation}

Equation (\ref{betun}) is not very convenient for the analysis of the wave
fronts because the parameter $\beta $ does not have a simple geometric
meaning related with the null surfaces. On the other hand, as we know, the
level surfaces of $S^{**}=u=const$ in Eq.(\ref{beta}) define the null
surfaces in which we are interested. Therefore, a sensible parameterization
will be the one that replaces the $\beta $ with the parameter $u$ defined by

\begin{equation}
u=\beta x^{a}l_{a}-H(\beta ,\zeta ,\bar{\zeta})=L(x^{a}l_{a},\zeta ,\bar{%
\zeta}).
\end{equation}

\noindent Constant values of $u$ label the characteristic surfaces
themselves and are different than $u_{B}=x^{a}l_{a}$. By changing the
parameter $\beta $ in favor of $u$ we are switching to the characteristic
description which provides a better framework to study the dynamics of the
wave fronts.\newline

\noindent
{\bf Remark 6}
{\it Note that $r\equiv \beta ^{-1}\eth\bar{{\eth}}S^{**}=\beta ^{-1}\eth\bar{{%
\eth}}u$ defines an affine parameter along the null geodesics that rule the
characteristic surfaces $u=const.$} \\

Instead of performing the transformation from the ``Cauchy
parameterization'' to the new set $(u,r,\zeta ,\bar{\zeta})$ we take a
shortcut, and start directly with the characteristic approach. Using the
notation of the previous sections for the characteristic problem the new
parameters are determined by the previous equations: $\ $ 
\begin{equation}
u=u(x^{a})=L(x^{a}l_{a},\zeta ,\bar{\zeta}),
\end{equation}
\noindent

\begin{eqnarray}
\dot{L}(x^{a}l_{a},\zeta ,\bar{\zeta})x^{a}m_{a}(\zeta ,\bar{\zeta})+{\ \eth}%
L(x^{a}l_{a},\zeta ,\bar{\zeta}) &=&0  \nonumber \\
\dot{L}(x^{a}l_{a},\zeta ,\bar{\zeta})x^{a}\bar{m}_{a}(\zeta ,\bar{\zeta})+%
\bar{{\eth}}L(x^{a}l_{a},\zeta ,\bar{\zeta}) &=&0,\noindent
\end{eqnarray}
\noindent and the new one defined by $r=\dot{L}^{-1}\eth\bar{{\eth}}S^{**}$
yielding

\begin{equation}
r=x^{a}(n_{a}-l_{a})+{\frac{\bar{{\eth}}\dot{L}}{\dot{L}}}x^{a}m_{a}+{\frac{{%
\eth}\dot{L}}{\dot{L}}}x^{a}\bar{m}_{a}+{\frac{\ddot{L}}{\dot{L}}}%
x^{a}x^{b}m_{a}\bar{m}_{b}+{\frac{{\eth}\bar{{\eth}}L}{\dot{L}}}.
\end{equation}

The coordinates $x^{a}$ can be written in terms of the four parameters $%
u,r,\zeta $ and $\bar{\zeta}$ as

\begin{equation}
x^{a}=u_{B}(l^{a}+n^{a})+(r+\bar{{\eth }}\Phi +\overline{\Phi }\dot{\Phi}
)l^{a}-\overline{\Phi }m^{a}-\Phi \bar{m}^{a},  \label{cambio}
\end{equation}

\noindent where

\begin{equation}
\Phi \equiv -{\frac{{\eth }L}{\dot{L}}},
\end{equation}

\noindent and the function $u_{B}\equiv x^{a}l_{a}$ is written in terms of
the parameters $u,\zeta $ and $\bar{\zeta}$ implicitly by $%
u=L(x^{a}l_{a},\zeta ,\bar{\zeta})$, i.e., $x^{a}l_{a}=L^{-1}(u,\zeta ,\bar{%
\zeta}).$ ($L^{-1}$ denotes the inverse \textit{function} of $L.$)

Treating Eq.(\ref{cambio}) as a coordinate transformation between the
natural coordinates associated with the solution, i.e., the $(u,\zeta ,\bar{%
\zeta},r),$ and the standard space-time coordinates $x^{a},$ the
transformation breaks down when its Jacobian vanishes. This is a
three-surface in the space-time; the caustic set associated with the
solution. \newline

After a lengthy calculation we find that this occurs when

\begin{equation}
J={\frac{\partial (t,x,y,z)}{\partial (u,r,\zeta ,\bar{\zeta})}}
=r^{2}-\sigma ^{0}\bar{\sigma}^{0}=0,  \label{5**}
\end{equation}

\noindent where

\begin{equation}
\sigma ^{0}={\eth}\Phi +\Phi \dot{\Phi}.  \label{6*}
\end{equation}

\noindent This is equivalent to Eq.(3.1) of reference\cite{new}\textbf{.}

There is a simple geometric interpretation of Eq.(\ref{5**}) and (\ref{6*});
the shear function $\sigma $ of the congruence of null geodesics that
generate the surfaces $u=constant,$ with the affine parameter $r,$ is given
by \cite{matristed}

\[
\sigma ={\frac{\sigma ^{0}}{r^{2}-\sigma ^{0}\bar{\sigma}^{0}}}. 
\]

\noindent Therefore, the vanishing of the Jacobian (\ref{5**}) implies that
the shear of the congruence diverges. We regain the expression defining
caustics from ref.\cite{gil} in the stationary case, namely,

\begin{equation}
r^{2}-{\eth }^{2}L{\ }\bar{{\eth }}^{2}L=0,
\end{equation}

\noindent since $\sigma ^{0}=\eth^{2}L$.

The form of the metric tensor in the new coordinates is

\begin{eqnarray}
ds^{2} &=&\eta _{ab}dx^{a}dx^{b}  \nonumber \\
&=&2{\frac{du}{\dot{L}}}\{dr+du({\frac{1+{\eth }\overline{\dot{\Phi}} +\Phi 
\overline{\ddot{\Phi}}}{\dot{L}}})  \nonumber \\
&+&d\bar{\zeta}({\frac{{\eth }\bar{\sigma}^{0}+\Phi \dot{\bar{\sigma}^{0} }-%
\dot{\Phi}\bar{\sigma}^{0}+\overline{\dot{\Phi}}{\ }r}{P}})+d\zeta ({\ \frac{%
\bar{{\eth }}\sigma ^{0}+\overline{\Phi }\dot{\sigma ^{0}}- \overline{\dot{
\Phi}}\sigma ^{0}+\dot{\Phi}{\ }r}{P}})\}  \nonumber \\
&-&{\frac{2r}{P^{2}}}(\sigma ^{0}d\zeta ^{2}+\bar{\sigma}^{0}d\bar{\zeta}
^{2})-2(r^{2}+\sigma ^{0}\bar{\sigma}^{0}){\frac{d\zeta d\bar{\zeta}}{P^{2}}}
,  \label{uu}
\end{eqnarray}

\noindent where $P=1+\zeta \bar{\zeta}$. This line element, corresponding to
shearing non-stationary null coordinates, defined by Eq.(\ref{cambio})
reduces to the one given in ref.\cite{gil} in the stationary regime. As
pointed out in this reference it might be of interest to use Eq.(\ref{uu})
as a background metric in linearized gravity for higher order perturbations
in problems where gravitational radiation is important.

\section{\noindent The Eikonal Equation in Asymptotically Flat Space-Times}

\qquad In a straightforward manner all our results can be generalized to the
case of arbitrary curved space-times, and the proofs of all the relation
above follow basically the same path. We will assume that there is \noindent
given a system of local coordinates $x^{a}$ in an arbitrary curved
space-time and a two parameter family (sphere's worth) of solutions of the
eikonal equation, i.e. $Z(x^{a},\zeta ,\bar{\zeta})$ such that

\begin{equation}
g^{ab}(x^{a})\partial _{a}Z(x^{a},\zeta ,\bar{\zeta})\partial
_{b}Z(x^{a},\zeta ,\bar{\zeta})=0  \label{eika}
\end{equation}
such that its (null) gradient sweeps out the light-cone at $x^{a}$ as ($%
\zeta ,\bar{\zeta})$ range over the sphere.

Such characteristic function $S=Z(x^{a},\zeta ,\bar{\zeta})$ are one of the
main variables of the Null Surface Formulation of General Relativity; they
contain all the conformal information of the space-time \cite{cacho}. In the
special case of asymptotically flat space-times $Z(x^{a},\zeta ,\bar{\zeta})$
can be interpreted either as the light cone cut of $\frak{{I}^{+}}$ of the
point with coordinates $x^{a}$, or as the past light cone of a point at $%
\frak{{I}^{+}}$ with coordinates ($u,\zeta $,$\bar{\zeta})$\cite{cacho}. 
\newline

We take the complete solution\textbf{\ $\beta Z(x^{a},\zeta ,\bar{\zeta})$, }%
and define, in an analogous manner to the flat-space construction,

\begin{equation}
u=S^{**}(x^{a},\beta ,\zeta ,\bar{\zeta})=\beta Z(x^{a},\zeta ,\bar{\zeta}%
)-H(\beta ,\zeta ,\bar{\zeta}).  \label{bota}
\end{equation}

\textbf{\noindent }On Eq\textbf{.}(\ref{bota}) we impose the conditions,
equivalent to (\ref{three}),(\ref{three*})\textbf{\ }and (\ref{three**}),
namely,

\textbf{
\begin{eqnarray}
\frac{\partial S^{**}}{\partial \beta } &=&Z(x^{a},\zeta ,\bar{\zeta})-\frac{%
\partial H(\beta ,\zeta ,\bar{\zeta})}{\partial \beta }=0 \\
\eth S^{**} &=&\beta \eth Z(x^{a},\zeta ,\bar{\zeta})-\eth H(\beta ,\zeta ,%
\bar{\zeta})=0 \\
\bar{\eth}S^{**} &=&\beta \bar{\eth}Z(x^{a},\zeta ,\bar{\zeta})-\bar{\eth}%
H(\beta ,\zeta ,\bar{\zeta})=0,
\end{eqnarray}
}

\noindent and solve for\textbf{\ ($\beta $, $\zeta $, $\bar{\zeta})$ }(as
earlier, always possible aside from lower dimensional (caustic) regions
which can be approached in a limiting fashion) in terms of the\textbf{\ $%
x^{a}$. }

When these are resubstituted into Eq.(\ref{bota}), \textbf{$S^{**}$} then
becomes a new solution of the eikonal equation\textbf{\ }since\textbf{\ }

\begin{equation}
\partial _{a}S^{**}=\beta \partial _{a}Z.  \label{flor*}
\end{equation}

As in the flat case, we can determine the arbitrary function\textbf{\ $%
H(\beta ,\zeta ,\bar{\zeta})$ }in terms of corresponding data given on a
Cauchy surface $\Sigma $\textbf{. }Suppose that we are given a coordinate
system\textbf{\ $(\tau ,x^{i})$ }such that $\tau =\tau _{0}$ corresponds to
our Cauchy surface, together with suitable Cauchy data\textbf{\ $%
S_{Cauchy}(x^{i})$ }on $\Sigma $\textbf{. }A needed generalization of the
relationship $\alpha _{i}=\partial S_{Cauchy}/\partial x^{i}$ from Sec.2 is

\textbf{
\begin{equation}
\frac{\partial S_{Cauchy}(x^{i})}{\partial x^{i}}-\frac{\beta \partial
Z(x^{i},\tau _{0},\zeta ,\bar{\zeta})}{\partial x^{i}}=0.
\end{equation}
}which is to be considered as three equations for the determination of 
\textbf{\ $x^{i}$ }in terms of\textbf{\ ($\beta $, $\zeta $, $\bar{\zeta}),$ 
}i.e., \textbf{$x^{i}=X^{i}(\beta ,\zeta ,\bar{\zeta})$. }When these are
inserted into Eq.(\ref{bota}) at \textbf{$\tau =\tau _{0}$ }we obtain

\textbf{
\[
H(\beta ,\zeta ,\bar{\zeta})=\beta Z(X^{i}(\beta ,\zeta ,\bar{\zeta}),\tau
_{0},\zeta ,\bar{\zeta})-S_{Cauchy}(X^{i}(\beta ,\zeta ,\bar{\zeta})). 
\]
}in analogy to the results of Sec.2.

\qquad The characteristic formulation from Sec.3 is even simpler. Starting
with any function $L(u_{B},\zeta ,\bar{\zeta})$ defined on $\frak{{I}^{+}}$,
we obtain a solution to the eikonal equation with the given characteristic
data by

\begin{equation}
u=S^{**}(x^{a},\zeta ,\bar{\zeta})=L(Z(x^{a},\zeta ,\bar{\zeta}),\zeta ,\bar{%
\zeta}),  \label{beto}
\end{equation}

\noindent where ($\zeta $,$\bar{\zeta})$ are functions of the coordinates $%
x^{a}$ such that the equivalent to Eq.(\ref{micon}) hold, i.e. when

\begin{eqnarray}
\eth S^{**} &=& \dot L \eth Z(x^{a},\zeta ,\bar{\zeta}) - \eth L(Z,\zeta,
\bar \zeta)=0 \\
\bar \eth S^{**} &=& \dot L \bar \eth Z(x^{a},\zeta ,\bar{\zeta}) - \bar
\eth L(Z,\zeta, \bar \zeta) = 0.
\end{eqnarray}

Again the relationship between $H(\beta ,\zeta ,\bar{\zeta})$ and $L(u,\zeta
,\bar{ \zeta})$ is given by the Legendre transformation

\begin{equation}
L(Z,\zeta ,\bar{\zeta})=Z\beta (Z,\zeta ,\bar{\zeta})-H(\beta (Z,\zeta ,\bar{%
\zeta}),\zeta ,\bar{\zeta}).  \label{G}
\end{equation}
with

\begin{eqnarray}
Z &=&\partial H/\partial \beta , \\
\beta &=&\partial L/\partial Z.  \nonumber
\end{eqnarray}

\section{Conclusion}

We have generalized the results of ref.\cite{gil} concerning solutions of
the flat-space eikonal equation. We saw two different means of giving data
and solving the eikonal equation: the Cauchy, and the characteristic
formulation. Each one leads to different methods. The two methods are
beautifully related by a Legendre transformation, Eqs.(\ref{4*}) and (\ref
{5*}). Moreover, all our results can be generalized to the case of curved
space-times. The characteristic formulation appears to be better for the
study of the dynamics of the wave fronts. By means of a suitable
parameterization we could describe the caustics in the wave-fronts, and find
a simple geometric interpretation in terms of the shear $\sigma $ of the
null congruence generating the wave fronts.

\section{Acknowledgments}

ETN thanks the NSF for support under grant \#PHY 97-22049 and AP thanks
FUNDACION YPF, Argentina, for its support.{\ }We extend our appreciation to
Simonetta Frittelli for many stimulating conversations.

\noindent


\begin{thebibliography}{9}
\bibitem{Arnold}  {V.I Arnold, \textit{Catastrophe Theory}, Springer Verlag,
Berlin, Heidelberg, NY, Tokyo, (1986). }

\bibitem{gil}  {Simonetta Frittelli, Ezra T. Newman and Gilberto
Silva-Ortigoza, }\textit{The Eikonal Equation in Flat Space; Null Surfaces
and Their Singularities}{, accepted for publication in JMP.}

\bibitem{landau}  {L. Landau and Lifschitz, \textit{Classical Mechanics},
Pergamon Press, Headington Hill Hall, Oxford, 4, 5 Fitzroy Sq. London, W.1,
(1960).}

\bibitem{new}  {B. Aronson, R. Lind, J. Messmer, and E.T. Newman, J. Math.
Phys., \textbf{12}, 2462, (1971).}

\bibitem{Arnold2}  {V.I Arnold, \textit{Mathematical Methods of Classical
Mechanics}, Springer Verlag, Berlin, Heidelberg, NY, Tokyo, (1980). }

\bibitem{matristed}  {R.W. Lind and E.T. Newman, J. Math. Phys., \textbf{15}%
, 1103, (1974). }

\bibitem{cacho}  {S. Fritelli, C. Kozameh and E.T. Newman, J. Math. Phys. 
\textbf{36}, 4984, (1995). }
\end{thebibliography}
\end{document}